\documentclass{sig-alternate-05-2015}

\usepackage{subfig}

\begin{document}

\CopyrightYear{2016}
\setcopyright{acmlicensed}
\conferenceinfo{ANRW '16,}{July 16 2016, Berlin, Germany}
\isbn{978-1-4503-4443-2/16/07}\acmPrice{\$15.00}
\doi{http://dx.doi.org/10.1145/2959424.2959438}

\title{Diurnal and Weekly Cycles in IPv6 Traffic}

\numberofauthors{1}
\author{
\alignauthor
	Stephen D. Strowes\\
	\affaddr{Yahoo Inc.}\\
	\email{sds@yahoo-inc.com}
}

\maketitle
\begin{abstract}

IPv6 activity is commonly reported as a fraction of network traffic per day.
Within this traffic, however, are daily and weekly characteristics,
driven by non-uniform IPv6 deployment across ISPs and regions.
This paper discusses some of the more apparent patterns we observe today.

\end{abstract}




\section{Introduction}

Global trends indicate that IPv6 accounts for approximately
10 -- 15\% of Internet traffic today, as shown in Fig.~\ref{f:global} and
corroborated by~\cite{google:ipv6} and \cite{isoc}.

The client choice between IPv4 and IPv6~\cite{rfc6724,rfc6555}
affects the network path characteristics observed by end hosts, given the
potential differences in peering arrangements, edge service availability, and configuration
mismatches (such as misconfigured host services, host or network firewalls).

IPv6 traffic patterns therefore affect considerations around load balancing,
hardware provisioning, network security, and peering; see also~\cite{czyz:2014:ipv6, livadariu:2016:stability, plonka:2015:ipv6}.
Thus, knowledge of traffic
patterns in networks and regions assists network planning.

\section{Global Traffic}

This analysis uses HTTP access logs collected on Yahoo's content
delivery network between April 26$^{th}$ 2016 and May 16$^{th}$ 2016
inclusive, excepting May 5$^{th}$, missing from this dataset. All timestamps are UTC.

\begin{figure}[t]
\includegraphics[width=\columnwidth]{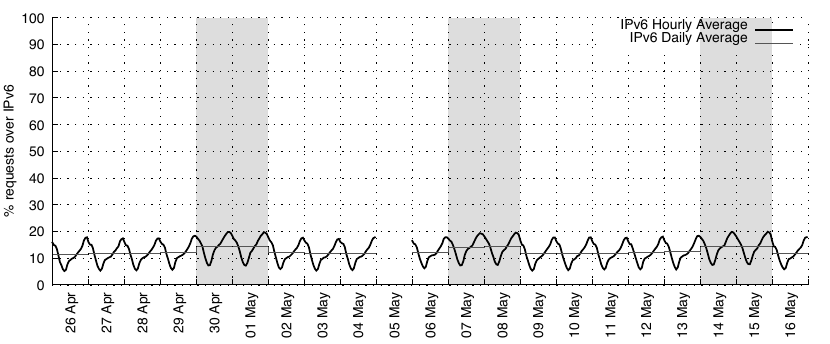}
\vspace{-1.6em}
\caption{Global hourly/daily averages of requests served over IPv6. Marked regions are weekends.}
\vspace{-0.9em}
\label{f:global}
\end{figure}

Fig.~\ref{f:global} shows hourly and daily global averages for the proportion
of requests served over IPv6. 
The daily cadence is clear, with notable differences between weekdays and weekends:
on weekdays, the IPv6 request ratio ranges from an average of 5.2\% between 06:00 -- 07:00UTC,
to 16.6\% by 22:00 -- 23:00UTC.
On weekends, the pattern is similar but with a higher minima of 7.4\% and maxima of 19.7\%.

The distinction between weekdays \emph{vs.} weekends
is echoed in the daily averages, with an average of 11.9\% requests
over IPv6 on weekdays, and 14.4\% on weekends. This variation is observed in
other published measurements~\cite{google:ipv6}. In this paper, we consider
contributing factors to those variations.

\subsection{Working-week Patterns}

\begin{figure}[t]
\includegraphics[width=\columnwidth]{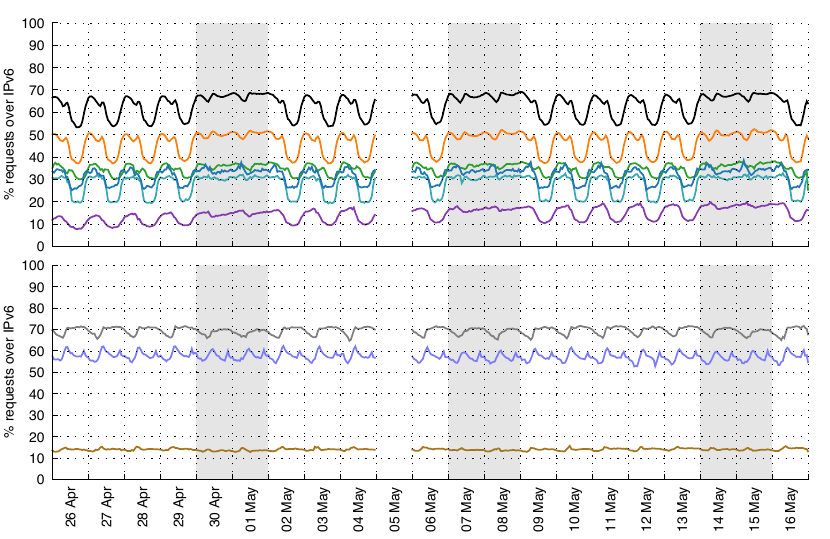}
\vspace{-1.5em}
\caption{US ISP traffic patterns; each line an ASN. \\ (a) Fixed-line providers. (b) Mobile providers.}
\vspace{-1em}
\label{f:isps-na}
\end{figure}

\begin{figure}[t]
  \includegraphics[width=\columnwidth]{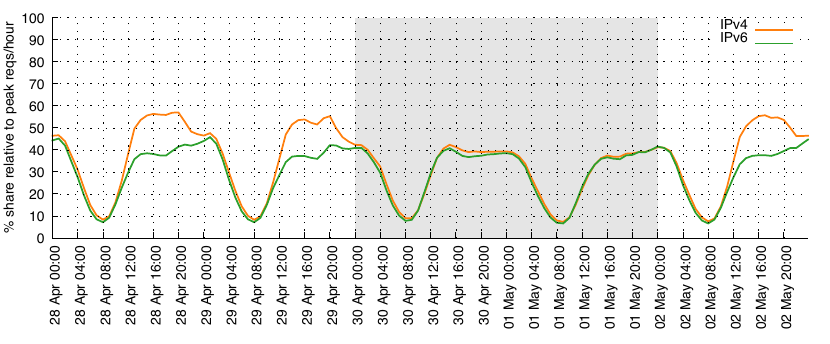}
  \vspace{-1.8em}
  \caption{Aggregate traffic for ASNs in Fig~\ref{f:isps-na}.}
  \vspace{-0.7em}
  \label{f:isps-traffic-na}
\end{figure}

\begin{figure}
\includegraphics[width=\columnwidth]{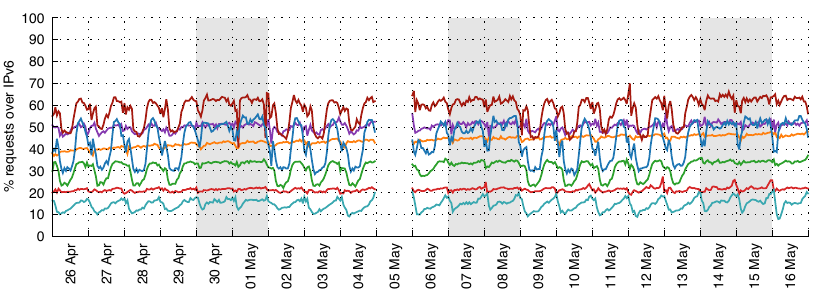}
\vspace{-1.6em}
\caption{European ISP traffic patterns; each line an ASN.}
\vspace{-0.9em}
\label{f:isps-eu}
\end{figure}

\begin{figure}[t]
  \includegraphics[width=\columnwidth]{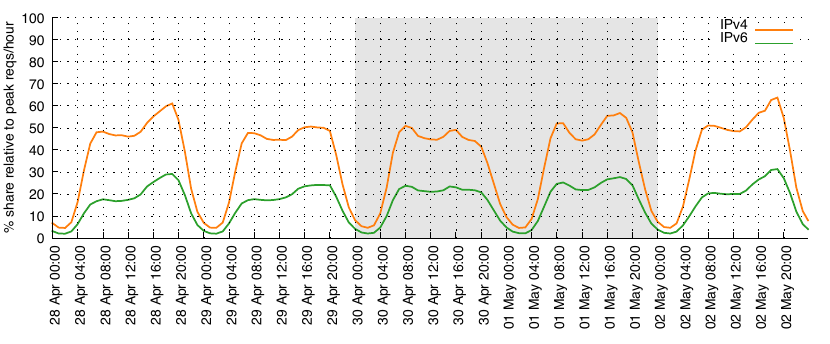}
  \vspace{-1.75em}
  \caption{Aggregate traffic for ASNs in Fig~\ref{f:isps-eu}.}
  \vspace{-0.5em}
  \label{f:isps-traffic-eu}
\end{figure}

Some ISPs have distinct access patterns that affect the IPv6 request
ratio through an ordinary week.
Fig.~\ref{f:isps-na} shows the daily share of requests from selected ASNs registered in the
US; these ASNs generate most of our US-based IPv6 traffic. ASNs for fixed-line service providers
are shown separately in Fig.~\ref{f:isps-na}a from mobile providers in Fig.~\ref{f:isps-na}b.

We observe
two behaviours: the fixed-line providers generate a distinct pattern during the working week
different from their weekend behaviour, while mobile providers are consistent throughout the week.
Regarding the fixed-line subscribers, to help determine whether the shift in request ratios occurs due to a decrease in requests over IPv6, an
increase in requests over IPv4, or a combination of both, request counts from the same set of ASNs are aggregated
in Fig.~\ref{f:isps-traffic-na}.
Noting that this plot shows absolute request counts as a fraction of the peak
observed in the aggregate dataset, we see requests dip on both protocols at
night, and peak during office hours, as expected.
Notably, the number of requests served over IPv6 is reasonably constant throughout, while there is a distinct
increase in requests over IPv4 during office hours. This implies that the lower IPv6 ratio is not an
obvious shift of requests from IPv6 to IPv4 but instead that additional traffic from office sites is
more likely to be IPv4-only, directly affecting the request ratios and thus the global
average reported on weekdays.

Fig.~\ref{f:isps-eu} shows the daily share of requests from selected ASNs registered in European
countries; these ASNs generate most of our EU-based IPv6 traffic. These are primarily fixed-line ISPs.
While a workday pattern is clear in some ASNs, there are various fixed-line providers that do not exhibit
the same daily pattern as the others; this may be due to the diverse nature of DSL providers in the
European market.
Fig.~\ref{f:isps-traffic-eu} shows the aggregate behaviour of these ASNs; largely, it appears that the 
absolute number of requests follows the same pattern between IPv4 and IPv6.

\subsection{Regional Traffic Patterns}

Regional variation in IPv6 deployment is clearly a contributing factor to
the daily variation in the share of requests over IPv6.
In order to determine \emph{where} IPv6 deployments are most active, we
can geo-locate requests and associate them with their likely time zone.

\begin{figure}[t]
\small
\subfloat{
  \includegraphics{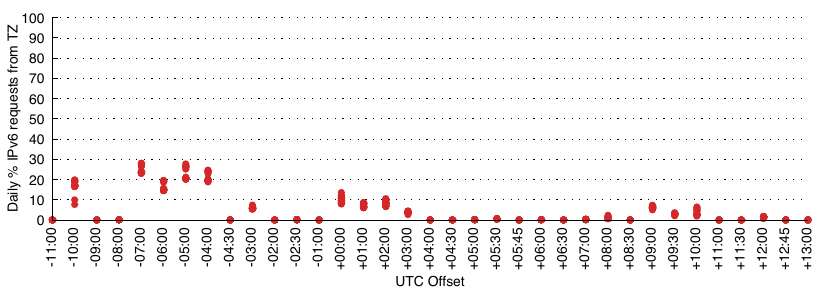}
  \label{f:tzs}
}
\vspace{-1.5em}

\subfloat{
  \includegraphics{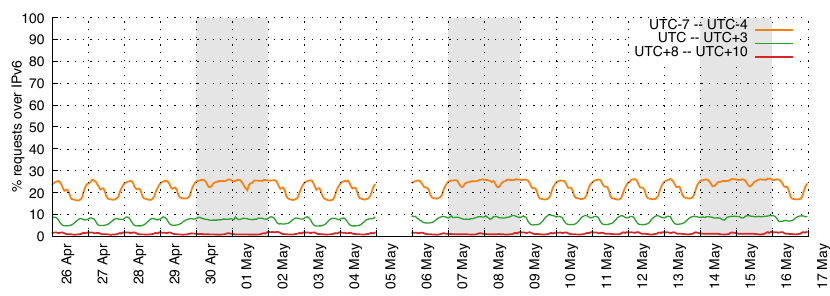}
  \label{f:timezones}
}
\vspace{-1em}
\caption{Traffic ratios across time zones. (a) All time zones observed. (b)
  Patterns for selected adjacent time zones.}
\vspace{-1em}
\label{f:regional}
\end{figure}

Fig.~\ref{f:tzs} shows the daily proportion of requests served over IPv6 from
each time zone observed. The notable clusters indicate the North American
countries, Brazil, various European countries (including Portugal, the UK,
France, Germany, Belgium, the Netherlands, the Czech Republic, Norway,
Lithuania, Estonia, Finland, Greece, and Romania), and some countries in the
Asia/Pacific region, such as Malaysia, Japan, and Australia.
Fig.~\ref{f:timezones} takes some of the active time zone clusters, and shows
their hourly averages separately.

The dominant regions are clearly North America and Europe.
Requests geo-located to North America are consistently
served more frequently over IPv6; commonly around 26\% of requests in the region between 00:00
to 06:00UTC on weekdays, \emph{i.e.}, peak evening hours, and a similar share through weekends
following the pattern observed in Fig.~\ref{f:isps-na}a. The IPv6 request share here consistently
falls to around 17\% between 12:00 and 20:00UTC.

In Europe, 9\% of requests tend to be made over IPv6 between 18:00 and 02:00UTC on weekdays
and most hours at weekends. The share of requests falls below 6.5\% between 06:00 and 14:00UTC.
As already noted in the European context, the correlation between increased IPv4 activity and core office hours
in some ASNs is not as strong as it is in the US.

\section{Conclusions}

The daily cadence shown in Fig.~\ref{f:global} is largely driven by
traffic from the US. The proportion of IPv6 requests is most pronounced in the US market,
and we see strong overlap between the quietest period in the US (08:00UTC)
and the lowest share of IPv6 in the global averages.
Also notable is the daily variation in IPv6 requests observed in many, but not all, fixed-line networks.
This is a key contributor to the 2.5\% difference in global IPv6 on weekdays \emph{vs.} weekends.

This paper presents a high-level overview of where and when IPv6 is active today. The protocol is clearly dominant in
some networks, and carries significant traffic in the US and the EU; aggregate
access patterns will continue to evolve as regions and networks
adopt the protocol.

\newpage

\bibliographystyle{abbrv}

\end{document}